\documentclass{article}
\usepackage{ijcai25}

\usepackage{times}
\usepackage{soul}
\usepackage{url}
\usepackage[hidelinks]{hyperref}
\usepackage[utf8]{inputenc}
\usepackage[small]{caption}
\usepackage{graphicx}
\usepackage{amsmath}
\usepackage{amsthm}
\usepackage{booktabs}
\usepackage{algorithm}
\usepackage[switch]{lineno}

\usepackage{algpseudocode}
\usepackage{cleveref}
\usepackage{xcolor} 
\usepackage{romannum}
\usepackage{amssymb}  % or \usepackage{amsfonts}
\usepackage{subcaption}
\usepackage{array} % required for text wrapping in tables
\usepackage{authblk}

\title{\textbf{Discovering "Words" in Music: Unsupervised Learning of Compositional Sparse Code for Symbolic Music}}

\author[1]{Tianle WANG}
\author[1]{Sirui Zhang}
\author[1]{Xinyi Tong}
\author[1]{Peiyang Yu}
\author[1]{Jishang Chen}
\author[1]{Liangke Zhao}
\author[5]{Xinpu Gao}
\author[4]{Yves Zhu}
\author[3]{Bo Zheng}
\author[2]{Duo Xu}
\author[4]{Yang Liu}
\author[2]{Xin Jin}
\author[1]{Feng Yu}
\author[2]{Songchun Zhu}

\affil[1]{Central Conservatory of Music, China, 
\affil[2]{Bigai, China, \texttt{\{mang33, jinxinbesti, s.c.zhu\}@...}}\texttt{22sa01@mail.ccom.edu.cn}}
\affil[3]{Alibaba Group, China, \texttt{\{tiezheng.gtz, bozheng\}@alibaba-inc.com}}
\affil[4]{Peking University, China, \texttt{\{xy.zhu, yangliu\}@pku.edu.cn}}
\affil[5]{Department of Industrial Engineering, Ajou University, Korea, \texttt{xinpu.gao@ajou.ac.kr}}

\begin{document}

\maketitle

\begin{abstract}
This paper presents an unsupervised machine learning algorithm that identifies recurring patterns--referred to as ``music-words''--from symbolic music data. These patterns are fundamental to musical structure and reflect the cognitive processes involved in composition. However, extracting these patterns remains challenging because of the inherent semantic ambiguity in musical interpretation. We formulate the task of music-word discovery as a statistical optimization problem and propose a two-stage Expectation-Maximization (EM)-based learning framework: 1. Developing a music-word dictionary; 2. Reconstructing the music data. When evaluated against human expert annotations, the algorithm achieved an Intersection over Union (IoU) score of 0.61. Our findings indicate that minimizing code length effectively addresses semantic ambiguity, suggesting that human optimization of encoding systems shapes musical semantics. This approach enables computers to extract ``basic building blocks'' from music data, facilitating structural analysis and sparse encoding. The method has two primary applications. First, in AI music, it supports downstream tasks such as music generation, classification, style transfer, and improvisation. Second, in musicology, it provides a tool for analyzing compositional patterns and offers insights into the principle of minimal encoding across diverse musical styles and composers. The source code and a demonstration page for our work are available in the public repository\footnote{\href{https://github.com/MusicWords/MusicWordsSrc.git}{MusicWords/MusicWordsSrc} }.
\end{abstract}

\section{Introduction}
\label{sec:intro}
\subsection{Music Is An Art of Repeating and Renewing}

From Bach concertos to pop tunes that today are 'ear-worm', the recurrence of patterns is a universal feature of music~\cite{margulis2013repetition,meyer2008emotion}. Computational aesthetic measures, such as information and complexity-based indicators~\cite{van2020order}, Zipf's law~\cite{zipf2016human}, and Gestalt-based metrics~\cite{lerdahl1996generative}, consistently demonstrated a strong association between the aesthetic appreciation of music and repetitive patterns. These patterns manifest at multiple structural levels, from small motifs to entire movements~\cite{margulis2013repetition,lerdahl1996generative}, often with subtle variations. In this paper, we refer to these repetitive units as ‘music-words.’

These templates function as fundamental building blocks for composition~\cite{margrave1968fundamentals} and are crucial in allowing listeners to perceive and recall musical structures ~\cite{deutsch2013psychology}. As a result, recognizing repetitive patterns is essential for machines to comprehend and generate musical content~\cite{muller2015fundamentals,conklin2003music,honing2006computational}.
    
\begin{figure}[t]
  \centering
   \includegraphics[width=1\linewidth]{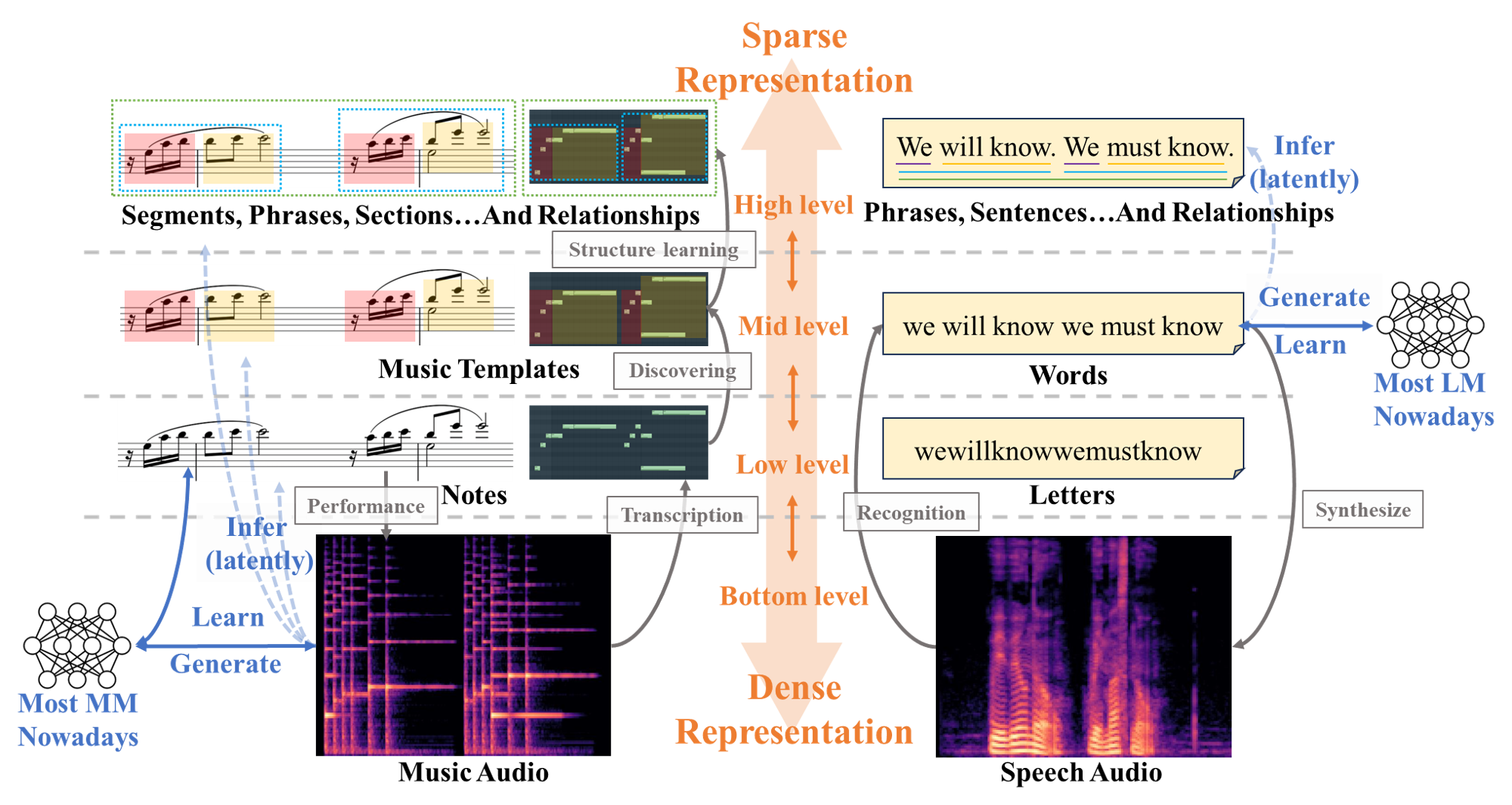}

   \caption{Different levels of Music and Natural Language Representation and AI tasks. 
   \textbf{Left}: Illustration of various levels of sparsity in music. 
   \textbf{Right}: Corresponding levels of representation in natural language and the workflow. 
   \textbf{From bottom to top}: transition from dense to sparse representation.}
   \label{fig:onecol}
\end{figure}
\subsection{The Absence of Middle Level Representation For Music}
However, to date, most mainstream music generation models and music information retrieval algorithms rely on audio data~\cite{dhariwal2020jukebox,engel2019gansynth,briot2020deep}, such as MusicGen~\cite{copet2023simple}, MusicLM~\cite{borsos2023musiclm}, and Riffussion~\cite{riffusion2023}, with a few exceptions, like MuseGan~\cite{dong2018musegan}, which operate on note-level data as shown in \cref{fig:onecol}. This challenge is analogous to attempting speech generation in natural language processing without a foundational concept of words or even  letters.

\begin{figure*}[t]
    \centering
    \includegraphics[width=1\linewidth]{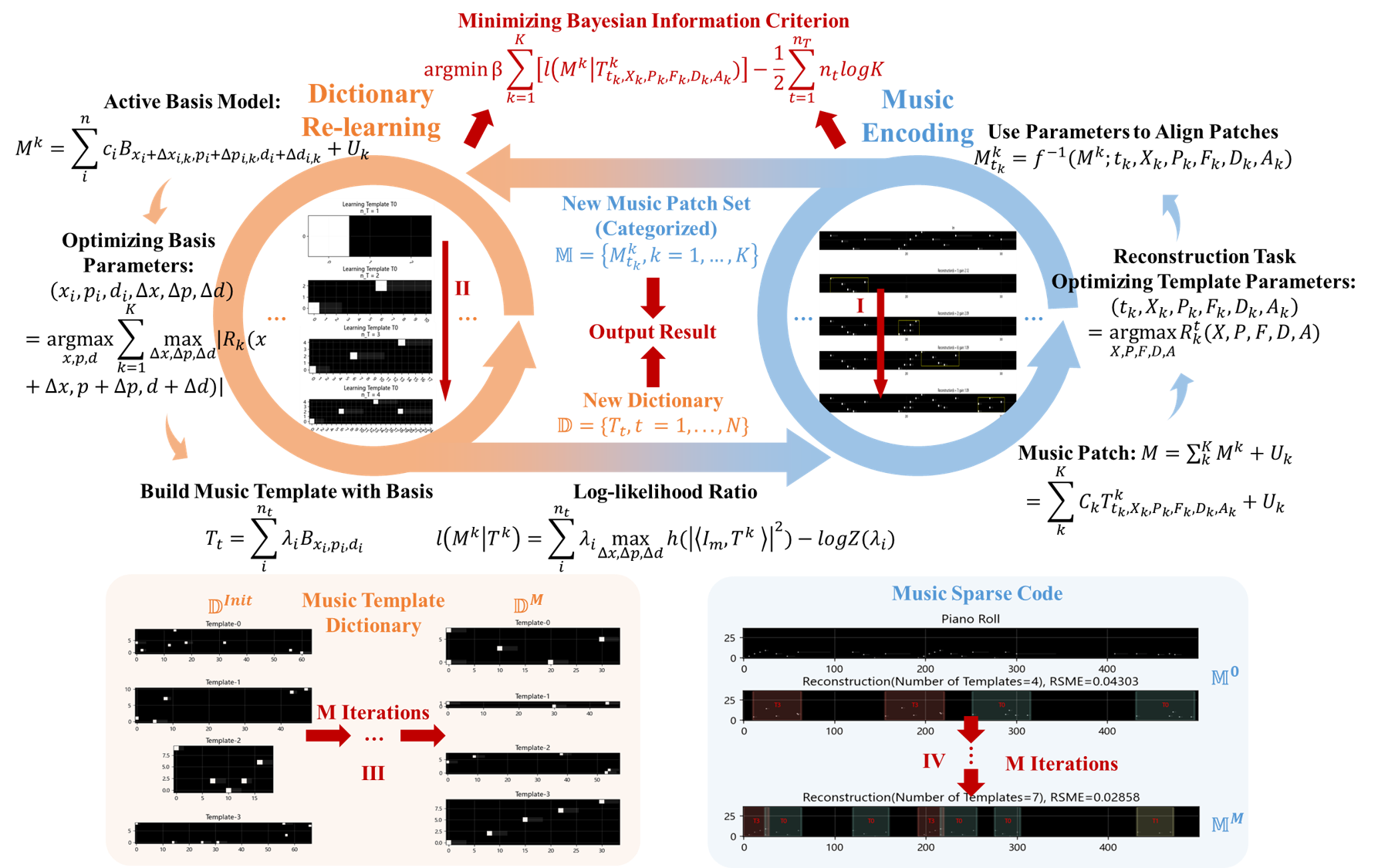}
    \caption{Overview of the compositional sparse coding model. 
    \textbf{\Romannum{1}}: music encoding. \textbf{\Romannum{2}}: dictionary re-learning. \textbf{\Romannum{3}}: music-word dictionary iteration. \textbf{\Romannum{4}}: music sparse code iteration}
    
    \label{fig:overview}
\end{figure*}

For humans, the creative process in music involves several stages, including composition, arrangement, performance, and mixing. While relying on large models and datasets to implicitly learn high-level structures across these stages is feasible, it remains a significant challenge. A prevalent expectation is that deep neural networks can autonomously derive music theory knowledge. However, this remains a significant hurdle, as powerful models tend to overfit to the dense, lower-level music data.

Just as word embedding techniques have become the foundation for almost every language model today, we propose the acquisition of a 'music dictionary/propose constructing a “music dictionary”. This intermediate representation provides a foundational framework for AI-driven music research, bridging the gap between low-level signal processing and high-level compositional analysis. By enabling exploration across multiple levels of musical abstraction, it facilitates a deeper understanding of music's fundamental structure.

\subsection{The Shortest Coding Length Principle}

\begin{figure}[t]
    \centering
    \includegraphics[width=1\linewidth]{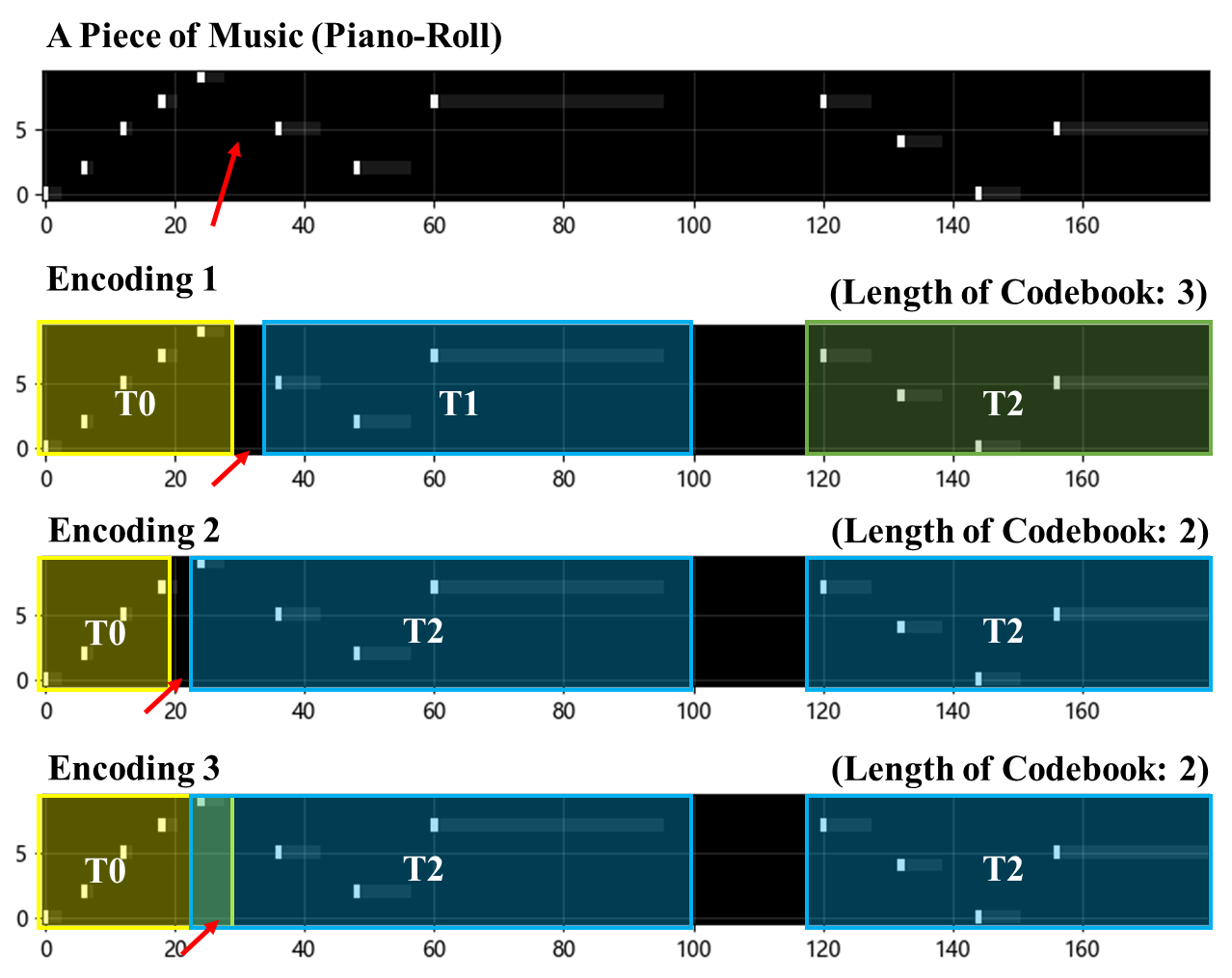}
    \caption{Examples of different encoding of music. }
    \label{fig:code len example}
\end{figure}

Despite its inherent appeal, the task of enabling machines to perform semantic segmentation on music—identifying and extracting recurring structural templates—remains highly challenging.
This difficulty arises from the inherent ambiguity and abstract nature of musical perception, which poses difficulties even for human experts. As illustrated in \cref{fig:code len example}, the first row shows a segment of music in a Piano-Roll format. If individuals were to segment the music based on this visual representation, they would likely make divisions at the positions marked by the red arrows, produces the segmentation labeled as \textbf{Encoding 1} as shown in the second row. However, upon listening to the entire piece, it becomes apparent that the segmentation in the third row is more reasonable. Moreover, certain interpretations consider that the fifth note in this segment is shared by both the preceding and succeeding music-words, as shown in the bottom row.
Upon examining the encodings, it is clear that the code lengths of the latter two, which are more widely accepted, are \textbf{Encoding 2} and \textbf{Encoding 3}, smaller than that of \textbf{Encoding 1}. This paper posits that the emergence of semantic subsets in music, language, and images is driven by the human desire for the shortest encoding length, which contributes to the simplicity and aesthetic quality of communication.

\subsection{Our Objectives and Approach}

As illustrated in Figure 2, our approach adopts an iterative two-stage Expectation–Maximization (EM)-style framework.
The first stage performs music encoding using a candidate music-word dictionary, while the second stage updates the dictionary based on the encoded results.
This alternating process is repeated until convergence, with the objective of jointly minimizing:
(1) the entropy of the learned music-words (achieving higher sparsity),
(2) the dictionary size (controlling model complexity), and
(3) the reconstruction error (ensuring faithful representation of the original music data).

\subsection{Contribution}
\begin{enumerate}

\item Introduced a probabilistic statistical model of ‘music-word discovery’ task with an optimization objective. Proposed an EM-style unsupervised learning algorithm that achieved an IoU accuracy of 0.61 compared to human expert annotations, without the use of any labeled data. Demonstrated the model’s potential for music structure analysis, highlighting how optimizing for sparsity reflects the shortest encoding principle, which may underlie human cognition of musical structure.

\item Defined a general form of sparse encoding representation for musical modal compositions and introduced a novel music similarity metric, BACC. Experimentally compared its performance with other similarity measures to assess its effectiveness.

\item Designed an efficient solution strategy that resulted in a 3860.8\% increase in training speed compared to the greedy strategy. Evaluated the performance of the music-word discovery algorithm across various scales, demonstrating its generalization ability and illustrating how scale variation aids in multi-layer music structure analysis.

\item Demonstrated the model’s ability to infer ‘word-like’ sparse codes for previously unseen music and analyze the frequency histogram of music-word usage, revealing that the distribution follows a power law. Framed music-word discovery as a template generation and matching problem, enabling the detection of cross-overlap instances. Conducted downstream testing for music classification and structure parsing, showing that mid-level structural information improves music information retrieval tasks.

\end{enumerate}

\section{Related Work} \label{sec:formatting}

\subsection{Musical Pattern Discovery}

Previous studies on repetitive patterns in music can be broadly classified into two categories: string-based methods~\cite{perkins2022musical,meredith2002algorithms}, which represent music as sequences of symbolic notes, and geometric methods~\cite{meredith2015music,buteau2001geometrical,meredith2002algorithms}, which treat music as entities in an n-dimensional space. This paper adopts the latter approach, offering two key advantages: 1) Additional dimensions capture more musical information beyond basic features like time, pitch, and duration, including dynamics, instrumentation, and techniques. 2) Insights from image pattern recognition can be leveraged.

However, this representation introduces challenges, such as harmonizing scales and information importance across dimensions.

It is important to distinguish between music structural segmentation~\cite{perkins2022musical,guan2018melodic,hernandez2020music,lattner2015probabilistic,bodily2021inferring}, which involves segmenting music along the time axis, and the task of discovering repeating patterns. The former is often applied, suited for non-overlapping, single-melody analysis. In contrast, this paper addresses the more complex task of pattern discovery.

While prior work in musical pattern discovery has focused on identifying recurring motifs~\cite{rolland2001motif,conklin1995multiple}, our study introduces differs by not only extracting music patterns~\cite{collins2011algorithmic} but also aiming to reconstruct the data in a non-redundant, complete manner~\cite{li2007music}, ensuring the learned music-word dictionary neither misses nor overrepresents patterns.

Computerized motif discovery methods often generate more results than human analysis~\cite{marsden2010musical}, leading to common filtering steps, such as discarding shorter patterns~\cite{cambouropoulos1998towards,collins2011pattern} or rejecting based on Gestalt principles~\cite{nieto2016discovering,lartillot2004musical}. In contrast, this study filters out less important templates while emphasizing shorter encoding lengths, with the final dictionary compiled by applying frequency statistics and sorting.

\subsection{ Active Basis }

The Active Basis Model, proposed by Wu et al.~\cite{wu2010learning}, is an image recognition algorithm that represents a class of images with a combination of a few basis functions. These functions can be fine-tuned to capture variations among similar images. This approach is suitable for identifying subtle differences in the same music-word across different positions, akin to its use in image recognition.

However, the model requires structurally aligned data, which is challenging in music, where templates appear at varying pitches, time positions, and transformations, making it impractical for learning music-words from diverse musical data.

\subsection{Compositional Sparse Code}
Hong et al.\cite{hong2014unsupervised}, building upon the Active Basis Model~\cite{wu2010learning}, established a higher level of hierarchical composition. They unified the overall parameters of the set of basis functions belonging to the same template category. This hierarchical structure allows the model to succinctly describe sparser mid-level structures in the data. Consequently, the problem of encoding the data is transformed into a two-layer parameter optimization problem: 1. Collective parameter optimization of multiple Templates composed of Basis functions; 2. Parameter optimization for each Basis within each template category.

By alternately solving these two layers of parameters, this method can unsupervisedly learn a few recurring image patterns from complex images to describe the original images. Inspired by this approach, this paper adopts a similar conceptual framework for music modeling and employs an improved solving algorithm to learn a music-word dictionary.

\section{Methods}

\subsection{Representation of Low-level Music Data}
Symbolic music is typically represented using musical notes as the smallest unit, with common representations including MIDI event sequences, sheet music notation, and piano roll matrices. We use a transformed Piano-Roll representation, where the horizontal axis represents time, the vertical axis represents pitch, and element values represent intensity~\cite{briot2017deep}. The details of this transformation process are provided in the \textcolor{blue}{\href{https://1drv.ms/b/c/6136db50f300f19c/EXTpJMekL31BqlZB1WXanf4BG1C_uyQR_x7SRZtQt8APgw?e=FmkVew}{supplementary materials}}.

\subsection{ Hierarchical Compositional Model}
\begin{figure}[h]
  \centering
  \begin{subfigure}{\linewidth}
      \centering
    \includegraphics[width=0.5\linewidth]{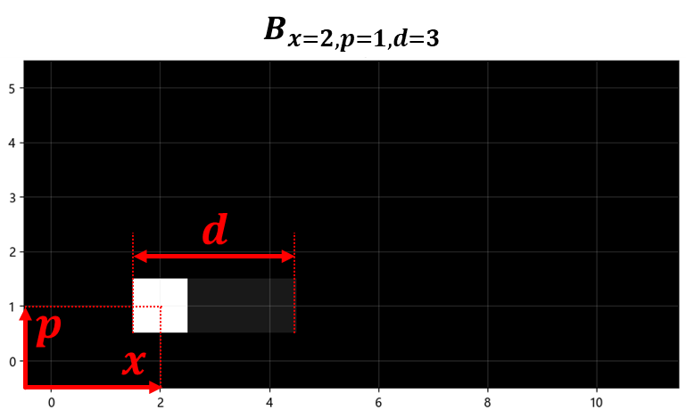}
    \caption{A music basis defined by 3 parameters: $p$: pitch; $x$: starting position; $d$: duration}
    \label{fig:basis}
  \end{subfigure}
  \hfill
  \begin{subfigure}{1\linewidth}
    \includegraphics[width=1\linewidth]{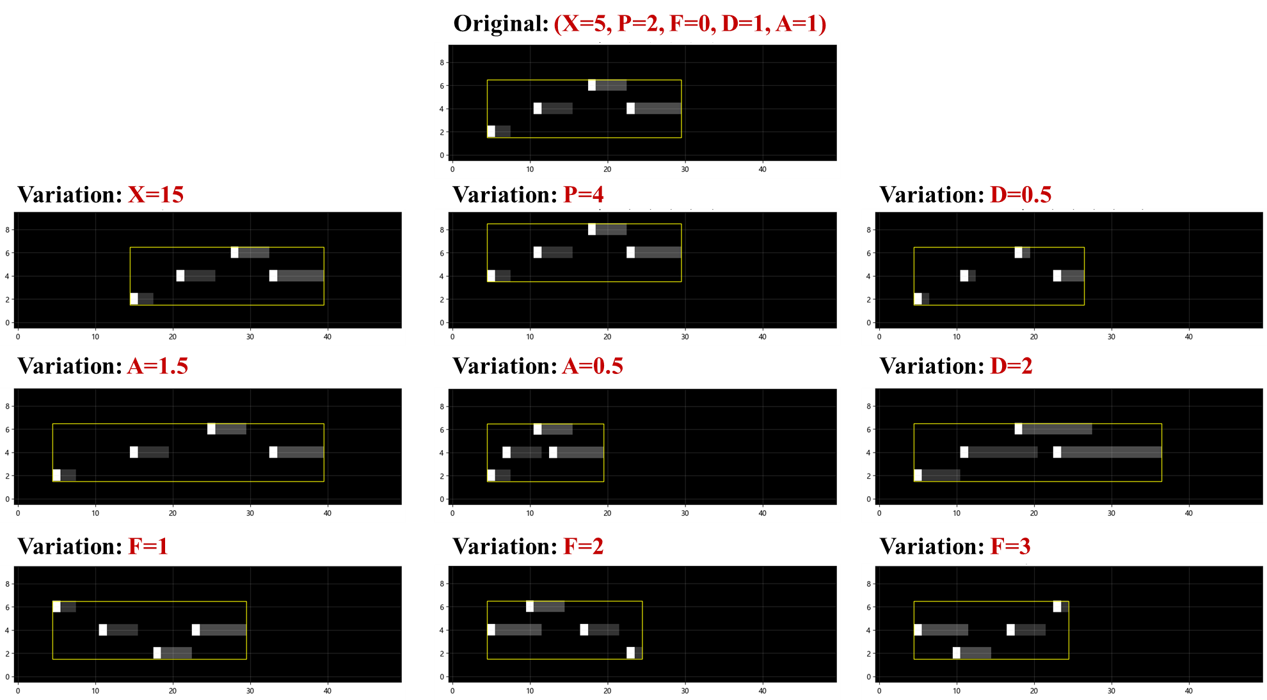}
    \caption{Deformed music-words with different geometrical parameters. (	t:type\ number\ of\ music-word\ ; 
	X:time\ offset\ ; 
	\ P:pitch\ offset; 
	F: flipping type;
	D: Duration scaling coefficient;
	A: Relative time distance coefficient)}
    \label{template}
  \end{subfigure}  \caption{
Variations of music basis and music-words}
  \label{fig:basis+template}
\end{figure}

As illustrated in \cref{fig:basis}, we treat individual musical notes as natural basis elements for our representation. Each basis, denoted as \(B_{x,p,d}\), has three parameters: pitch $p$, starting position $x$, and duration $d$. Therefore, a category of music-word \textit{$T_t$} can be uniformly represented as follows:

\begin{equation}
T_t=\sum_{i}^{n_t}{\lambda_iB}_{x_i,p_i,d_i}
  \label{}
\end{equation}
Consider a set of music patches  
\(\mathbb{M}=\left\{M_k^0,k=1,\ldots,K\right\}\) where each music patch has a similar basic shape, but with minor differences in some basis.

The form of the active basis model is as follows:
\begin{equation*}
M_k=\sum_{i=1}^{n}{c_iB_{x_i+\mathrm{\Delta}x_{i,\ \ k},\ p_i+\mathrm{\Delta}p_{i,\ \ k},\ d_i+\mathrm{\Delta}d_{i,\ \ k}}+U_k}
\end{equation*}

where \(\mathrm{\Delta}x_{i,k}, \mathrm{\Delta}p_{i,k}, \mathrm{\Delta}d_{i,k}\) denote the adaptive fine-tuning quantities for the starting time position, pitch, and duration of the basis concerning that music patch instance (default settings: \(\mathrm{\Delta}x_{i,k}\in\left[-2,2\right], \mathrm{\Delta}p_{i,k}\in\left[-2,2\right], \mathrm{\Delta}d_{i,k}\in\left[-5,5\right]\ \))

We can apply overall geometric transformations to a music-word \(T_t\) formed by multiple notes, such as flipping, scaling, translation, etc., which are common variations in music composition. The geometric properties of the transformed music-word \(T_{t,X,P,F,D,A}\) are defined as follows: 
\begin{itemize}
                \item X:time\ offset
                \item P:pitch\ offset
                \item F: flipping type (0-original; 1- horizontal flip; 2- vertical flip; 3-diagonal flip)
                \item D: Duration scaling coefficient
                \item A: Relative time distance coefficient
\end{itemize}

Assuming a given music-word dictionary \(\mathbb{D}=\{{T_t,t\ =1,\ ...,\ N}\}\), we sequentially select music-words from the desired categories, transform them, and use them for encoding \(\textit{\textbf{M}}\) 
\begin{equation*}
M=\sum_{k}^{K}{C_kT_{t_k,X_k,P_k,F_k,D_k,A_k}^k+U_k}
\end{equation*}
\subsection{ Similarity Measure }
The paper experimented with three different similarity metrics to approximate the matching degree between music-words and music patches, including:
\begin{enumerate}
    \item  Zero-normalized cross-correlation (ZNCC)

\begin{equation}
\begin{aligned}
\small
&\mathbf{ZNCC}\left(\mathbf{X},\mathbf{P}\right)\\
&=\frac{1}{n}\sum_{x,p}\frac{\left(\mathbf{M}\left(X+x,P+p\right)-\mu_\mathbf{M}\right)\left(\mathbf{T}\left(x,p\right)-\mu_\mathbf{T}\right)}{\sigma_\mathbf{M}\sigma_\mathbf{T}}
\end{aligned}
\end{equation}

    \item Root mean square error (RMSE)

\begin{equation}
\small
\mathbf{RMSE}\left(\mathbf{X},\mathbf{P}\right)=\sqrt{\frac{1}{n}\sum_{x,p}\left(\mathbf{M}\left(\mathbf{X}+x,\mathbf{P}+p\right)-\mathbf{T}\left(x,p\right)\right)^2}
\end{equation}

    \item Basis-average cross-correlation (BACC)

\end{enumerate}
\begin{equation}
\small
\mathbf{BACC}\left(\mathbf{X},\mathbf{P}\right)=\frac{1}{n^{(T)}+n^{(M)}}\sum_{x,p} \mathbf{M}\left(\mathbf{X}+x, \mathbf{P}+p\right) \mathbf{T}\left(x,p\right)
\end{equation}

Here, \(n^{\textit{\textbf{T}}}\)and \(n^{\textit{\textbf{M}}}\)represent the number of basis in \textbf{T} and the number of basis in \(\textit{\textbf{M}}\) within the sliding window, respectively. 

The number of non-zero elements in a music Piano-Roll is generally lower than the number of non-zero pixels in a natural image. Therefore, the normalized way of ZNCC may not be suitable for symbolic music data. In this paper, a modified version of ZNCC for Piano-Roll is proposed, which uses the number of notes contained in the music patch and music-word as normalization factors instead of the number of pixels and mean values.

We expect a good similarity metric to:
\begin{enumerate}
    \item Have high response values for slightly different but similar music patterns.
    \item Yield a lower response when the number of notes in the music patch is \textbf{greater} than the number of basis in the music-word.
    \item Yield a lower response when the number of notes in the music patch is \textbf{fewer} than the number of basis in the music-word.
\end{enumerate}
After experimentation, we ultimately choose BACC as the default metric.

\subsection{Statistical Modeling}
The distribution of \(M_k\) given the deformed music-word \(p\left(M_k\right|T_k)\) can be viewed as the migration from the original distribution \(q(M_k)\)

\[p\left(M_k\right|T_k)=q(M_k)\prod_{i=1}^{n}\frac{p_i\left(c_{k,i}\right)}{q\left(c_{k,i}\right)}\]
We adopt the exponential family model to model the distribution of \(c\) : 

\[p\left(c;\lambda\right)=\frac{1}{Z\left(\lambda\right)}e^{\lambda h\left(\left|c\right|^2\right)}q\left(c\right)\]
\(h(r)\) is the activation function used to suppress overly large responses during the active basis process. 
\[h\left(r\right)=\xi(\frac{2}{1+e^{-\frac{2r}{\xi}}}-1)\]
The normalization constant can be obtained through integration, i.e., the expected value of the exponential term under the \(q(c)\)  distribution:
\[Z\left(\lambda\right)=\int{e^{\lambda h\left(r\right)}q\left(c\right)dc}=E_q\left[e^{\lambda h\left(r\right)}\right]\]
Thus, we can evaluate the proximity of the current model distribution to the reference model distribution through Kullback-Leibler divergence, calculating the Log-likelihood ratio: 

\begin{equation}
\begin{aligned} 
    &l\left(I_k\middle| T,T_k,\Lambda\right)=\sum_{k=1}^{K}{log\frac{p\left(M_k\middle| B_k\right)}{q\left(I_k\right)}}\\
    &=\sum_{k=1}^{K}\sum_{i=1}^{n}\lambda_ih(Mk,B_{xi+\Delta xi,  k, pi+\Delta pi,  k, di+\Delta di, k2})-logZ(\lambda_i)
\end{aligned} 
\label{eq:likelihood}
\end{equation}

\subsection{Overview}
To solve the optimization problem \cref{eq:likelihood}, we propose a two-stage alternating algorithm, as shown in \cref{fig:overview}. This framework alternately solves two tasks:

\begin{enumerate}
    \item \textbf{Music Encoding}: Using a given music-word dictionary, the goal is to determine the music-word parameters and encode the music data with the objective of the reconstruction loss is minimized.
    \item \textbf{Dictionary Re-learning}: Using the encoding results from the previous step, salient music patches are extracted to form a temporary training set for dictionary updating. The active basis model is then applied to re-estimate a representative music-word for each category, thereby optimizing the alignment between the dictionary entries and their corresponding instances.

\end{enumerate}

The specific algorithm can be found in the pseudo code in the \textcolor{blue}{\href{https://1drv.ms/b/c/6136db50f300f19c/EXTpJMekL31BqlZB1WXanf4BG1C_uyQR_x7SRZtQt8APgw?e=FmkVew}{supplementary materials}}.

\subsection{Efficient Strategy }
Greedy strategy guarantees the optimal solution at each iteration; however, it is computationally expensive, making it challenging for the algorithm to scale to larger music datasets. To address this limitation, we introduce a set of computational enhancements designed to accelerate both the music encoding and dictionary re-learning stages of the EM-style framework. Although these improvements forgo the guarantee of the optimal solution, they significantly increase the model's efficiency. The specific strategies—including random-cropping initialization, Gaussian-diffuse matching, and incremental learning—are detailed in the \textcolor{blue}{\href{https://1drv.ms/b/c/6136db50f300f19c/EXTpJMekL31BqlZB1WXanf4BG1C_uyQR_x7SRZtQt8APgw?e=FmkVew}{supplementary materials}} to facilitate reproducibility.

\section{Experiments}
\subsection{Implementation Details}
We set the MIDI batch processing time length to 46,080 ticks (or 96 beats for a Tick\_per\_beat of 480) and down-sample the temporal resolution to Tick\_per\_beat = 12 to accelerate computation. The dictionary is initialized with $n_t = 4$ entries, with an initial music-word size of $10 times 20$. The music-word scaling factors $X, P, D, A \in   (0.5, 0.8, 1, 1.2, 1.4)$, and the basis count penalty factor is set to $\gamma = 1$. We use a Gaussian kernel with $size = 5$ (preferably odd), Gaussian distribution parameter $\sigma = 2$, uniqueness filtering threshold $u = 0.4$, and significance threshold $s = 0.5$. The model is trained on 30 popular songs over 10 epochs using an NVIDIA GeForce RTX 3070 with 8.0 GB of memory.

\subsection{Training and Evaluation}
\begin{figure}[t]
    \centering
    \includegraphics[width=1\linewidth]{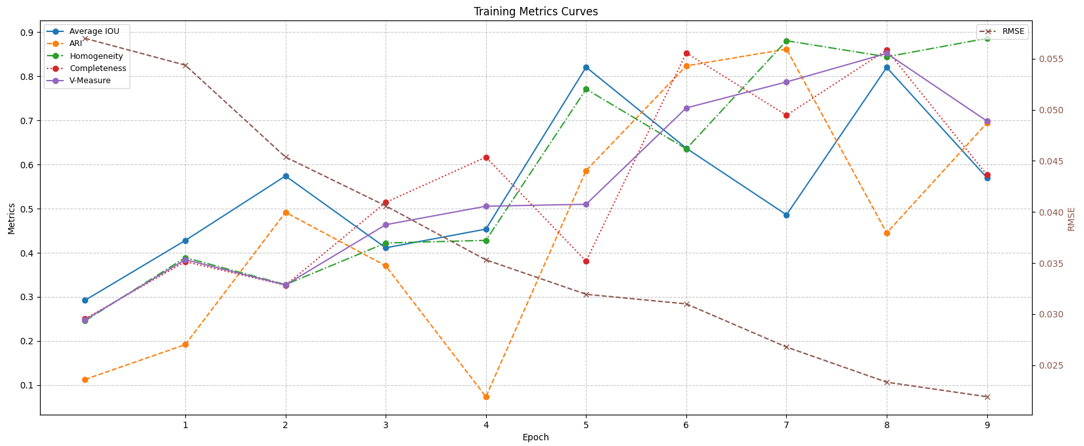}
    \caption{Sparse encoding performance over 10 epochs of training. }
    \label{fig:train curve}
\end{figure}

 \cref{fig:train curve} presents the performance metrics collected from 30 annotated music samples over 10 training epochs, including reconstruction error, Intersection-over-Union (IoU), homogeneity, completeness, and Adjusted Rand Index (ARI). The chart indicates the following:

\begin{table*}[htbp] 
  \centering
  \begin{tabular}{@{}ll|lllllll}
    \toprule
    \textbf{Strategy}&\textbf{Similarity}&\textbf{IOU}\ \textuparrow& \textbf{Homogeneity}\ \textuparrow& \textbf{Completeness}\ \textuparrow& \textbf{V-Measure}\ \textuparrow& \textbf{RMSE} \ \textdownarrow& \textbf{ARI}\ \textuparrow&\textbf{Sec/Epoch} \ \textdownarrow\\
    \midrule
    {Greedy}&{\textbf{BACC}}&\textbf{.611}& .720& .541& .618& \textbf{.019}&  .672&4951 \\
 {\textbf{Efficient}} &{\textbf{BACC}}&.570& \textbf{.886}& .577& \textbf{.699}& .022&  .\textbf{695}&\textbf{125}\\
 \textbf{Efficient}  &ZNCC&.562& .632& \textbf{.664}& .648& .830&  .620&149 \\
    \textbf{Efficient}  &RMSE&.523& .615& .409& .491& .020&  .661&151\\
    \bottomrule
  \end{tabular}
  \caption{Ablation study of different resolving strategies and music similarity measures.}
  \label{tab:example }
\end{table*}
Although the 'music segmentation' task differs in task definition from the 'music pattern discovery' task focused on in this paper, the proposed method can still be applied for prediction in the music segmentation task with appropriately set parameters. Reed Perkins and Dan Ventura\cite{perkins2022musical} explored the performance of various semantic reasoning algorithms in the phrase segmentation task. This paper adopts the same evaluation approach, conducting zero-shot training and prediction on the same labeled dataset\footnote{Essen dataset: \href{http://www.esac-data.org/}{http://www.esac-data.org/}}\footnote{JKUPDD dataset: \href{https://www.tomcollinsresearch.net/mirex-pattern-discove ry-task.html}{https://www.tomcollinsresearch.net/mirex-pattern-discove ry-task.html}}\footnote{Hymns dataset: \href{https://github.com/reedper kins/hymns-dataset}{https://github.com/reedper kins/hymns-dataset}}. The resulting F1-scores, summarized in \cref{tab:compare}, demonstrate that our method (Efficient+BACC) consistently outperforms syntactic baselines across all three datasets.
\begin{table}[htbp] 
  \centering
  \begin{tabular}{>{\raggedright\arraybackslash}p{0.45\linewidth}|lll}
    \toprule
    &Essen &Hymns &JKU\\
    \midrule
    Gramatic (LongestFirst+VCL-best)&0.35& 0.5&0.58\\
 \textbf{Ours (Efficient+BACC)}&\textbf{0.42}& \textbf{0.56}&\textbf{0.63}\\
    \bottomrule
  \end{tabular}
  \caption{A Comparison of F1 Scores for Music Segmentation Across Three Datasets Using Syntactic Methods.  }
  \label{tab:compare}
\end{table}

\textbf{Results Analysis} 
\begin{enumerate}
\item Although the efficient algorithm does not guarantee improvement at every step and exhibits some local oscillations, it generally enhances accuracy. 
\item The reconstruction error steadily decreases, a trend we attribute to the music-word regeneration strategy employed. 
\item Although the mean IoU fluctuates across epochs, it converges from approximately 0.3 to 0.6, as shown in \cref{fig:train curve}(b). This overall improvement indicates a stronger alignment between the discovered music-words and human annotations. The observed variability stems from the model's relaxed boundary constraints—a design choice that affords flexibility but occasionally incorporates marginal regions during pattern detection.
\item The Adjusted Rand Index (ARI) measures the similarity between the predicted and true labels, adjusted for chance. A value of 0.581 indicates moderate to strong agreement between the predicted and true cluster assignments, though there is still room for improvement. 
\item Homogeneity measures how well each cluster contains only members of a single class. A value of 0.89 is relatively high, suggesting that the algorithm is effective in ensuring that the samples within each predicted cluster mostly belong to the same true class. This indicates that the clustering model is fairly accurate in grouping similar instances together. 
\item The completeness score of 0.58 suggests that while some true classes are captured well, others may be fragmented across multiple clusters, leading to incomplete groupings. Ideally, completeness would be closer to 1, indicating that all instances from a true class are assigned to the same cluster. 
\item A value of 0.7 indicates a good overall balance between homogeneity and completeness, although the lower completeness score suggests that there is room for improvement in fully capturing the relationships between all classes.
\end{enumerate}

\subsection{Coding Results}
\begin{figure}[t]
    \centering
    \includegraphics[width=1\linewidth]{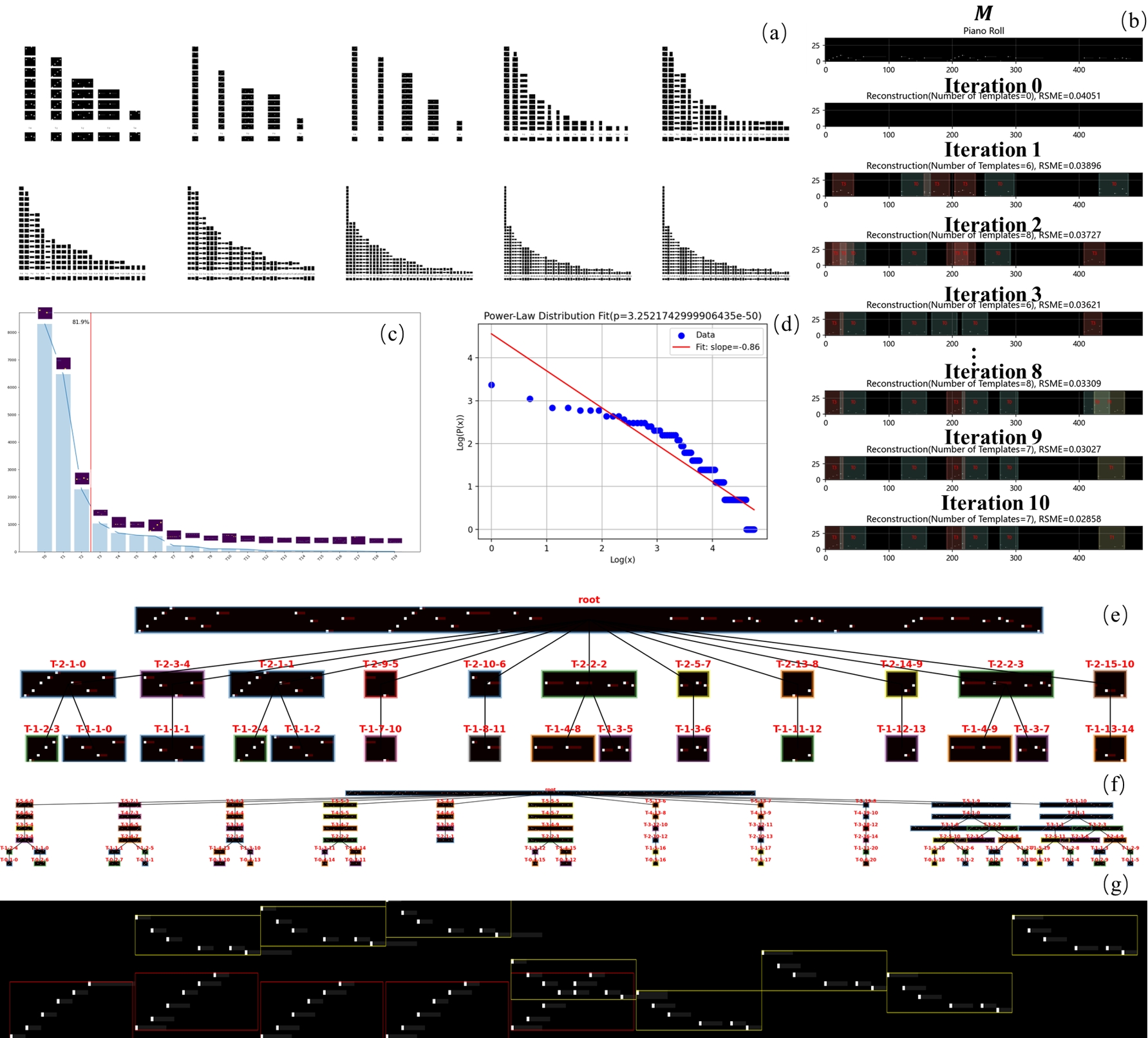}
    \caption{Sparse coding results. \textbf{(a)} Iteration of the dictionary in the incremental learning process. \textbf{(b)} Iteration of coding result of a music segment over 10 epochs. \textbf{(c)} Frequency of 20 most occurred music-words, the red line indicates that the top 3 most frequent music-words account for 81.9\% of all music-words. \textbf{(d)} Power-law distribution fit. \textbf{(e)} Example of 3-layer hierarchical sparse code of a music segment using multi-size sparse code model. music-words labeled as T-{layer}-{class index}-{instance index}. \textbf{(f)} Example of 7-layer hierarchical sparse code of longer music. \textbf{(g)} Example of successfully sparse coding for polyphonic music with overlapping notes.}
    \label{fig:coding results}
\end{figure}

\textbf{Dictionary}
The dictionary capacity grows adaptively with the volume of input data, exhibiting a progressive increase in basis functions as shown in \cref{fig:coding results}(a). In each inner loop of the re-learning phase, the dictionary is rebuilt from scratch and grows by adding basis functions, as shown in process \textbf{II} of \cref{fig:overview}. In the larger loop, the dictionary’s contents are gradually refined, and its length increases, corresponding to process \textbf{III} of \cref{fig:overview}. Meanwhile, the exponent $ \alpha$ of the power-law distribution gradually increases, indicating that the sparsity of the distribution is increasing, as illustrated in \cref{fig:coding results}(c) and (d), with $\alpha=0.86(p=3.25 \times 10^{-50} <0.05)$.

\textbf{Music Sparse Code}
The sparse coding results of the music data, shown in process \textbf{I} of \cref{fig:overview}, progressively approximate the original music data as each music-word is added one by one. In the larger loop, after $M$ alternating updates, the encoding results lead to a continuous decrease in the reconstruction error, \cref{fig:coding results}(b).

\textbf{Special Case — Overlapping}
Because the model frames music-word extraction as a coding optimization rather than a segmentation task, it effectively handles complex cases involving overlapping or interwoven patterns in polyphonic music (\cref{fig:coding results}g). 

\subsection{Structure Learning}
By gradually increasing the maximum duration generated by the music-word, we can observe the repetitive structures of music at various hierarchical levels such as phrases, sections, measures, and movements. The overlapping structures at different levels produce multi-level parse trees, as shown in \cref{fig:coding results}(e), (f).

\subsection{Music-Word Application Test: Music Classification}
To validate the practical significance of sparse coding for downstream tasks, we use music-word information parsed from MIDI files as additional high-level features in a music classification model. The goal is to assess whether incorporating these features improves classification performance (see \textcolor{blue}{\href{https://1drv.ms/b/c/6136db50f300f19c/EXTpJMekL31BqlZB1WXanf4BG1C_uyQR_x7SRZtQt8APgw?e=FmkVew}{supplementary materials}} for experimental setup).

\begin{table}[htbp] 
  \centering
  \begin{tabular}{@{}ll|ll}
    \toprule
    \textbf{Represent.}&\textbf{Feats.}&\textbf{Composer}\ \textuparrow& \textbf{Difficulty}\ \textuparrow\\
    \midrule
    Matrix&basic&0.572& \textbf{0.420}\\
 Matrix&advanced&0.618& 0.415\\
 Matrix&\textbf{music-words}&\textbf{0.620}& 0.413\\
    Sequence&basic&0.447& \textbf{0.368}\\
 Sequence& advanced& 0.393&0.349\\
 Sequence& \textbf{music-words}& \textbf{0.418}&0.342\\
 
    \bottomrule
  \end{tabular}
  \caption{Accuracy of three identification tasks on the ASAP dataset, with basic features, higher-level features, and music-words.}
  \label{tab:classi }
\end{table}

Compared to other simple features, by using sparse coding as a high-level feature, the model showed a slight improvement in accuracy for the composer prediction task. This improvement is likely attributable to the common music word features capturing the habits of different composers, where specific chord textures and performance habits become their 'personal style signatures.' For example, when humans hear a series of rapid, tense note progressions, they would immediately guess it's a Chopin piano piece. However, for difficulty prediction, compared to other computable features, music words may not have brought a significant performance improvement. This limitation likely arises from the fact that the difficulty labels in the ASAP dataset emphasize local physical factors (e.g., finger span) rather than global structural features, suggesting that incorporating fine-grained biomechanical descriptors may further enhance prediction performance

\section{Discussion and Future Work}
This paper proposes a statistical probabilistic model for music repetitive fragment discovery, inspired by computer vision, and solves it using an EM iterative unsupervised learning algorithm. The study explores different solution strategies and similarity measures, ultimately achieving an accurate music word discovery model via sparse coding. It also conducts statistical distribution analysis of the results, highlighting the potential application value in musicology, human-machine co-creation, and music perception. Experiments involving multi-scale analysis and polyphonic music demonstrate that the proposed method achieves superior generalizability compared with existing approaches. Moreover, preliminary music classification experiments confirm the value of mid-level music representations for downstream tasks such as composer identification.

The research has several areas for improvement and future work:

\begin{enumerate}
    \item \textbf{Accuracy improvement}: The experiments mainly validate the theory, with room for better parameter tuning and algorithm enhancement.
    \item \textbf{Adding note dimensions}: The current 2-D matrix representation could be expanded to include timbre, dynamics, and pedal usage, though stability and convergence may vary.
    \item \textbf{Word vector embedding}: The proposed music words could be vectorized using self-supervised pretraining, enabling richer semantic encoding and better learning of music-specific properties.
    \item \textbf{Music generation}: This method helps extract a musician’s music lexicon, potentially enabling AI improvisation, composition, and human-machine collaboration.
    \item \textbf{Efficient solving methods}: While solution strategies have improved, iterative solving still takes time. Deep learning models could be used to propose music words and optimize for real-time applications.
    \item \textbf{Support for audio input}: The method could be extended to extract music words directly from audio chromagrams, increasing its scope and value for tasks like source separation and audio-to-score transcription.
\end{enumerate}

\bibliographystyle{named}
\bibliography{ref}

\begin{thebibliography}{}

\bibitem[\protect\citeauthoryear{Agostinelli \bgroup \em et al.\egroup
  }{2023}]{borsos2023musiclm}
Andrea Agostinelli, Timo~I Denk, Zal{\'a}n Borsos, Jesse Engel, Mauro Verzetti,
  Antoine Caillon, Qingqing Huang, Aren Jansen, Adam Roberts, Marco
  Tagliasacchi, et~al.
\newblock Musiclm: Generating music from text.
\newblock {\em arXiv preprint arXiv:2301.11325}, 2023.

\bibitem[\protect\citeauthoryear{Bodily and
  Ventura}{2021}]{bodily2021inferring}
Paul~Mark Bodily and Dan Ventura.
\newblock Inferring structural constraints in musical sequences via multiple
  self-alignment.
\newblock In {\em Proceedings of the Annual Meeting of the Cognitive Science
  Society}, volume~43, 2021.

\bibitem[\protect\citeauthoryear{Briot \bgroup \em et al.\egroup
  }{2017a}]{briot2020deep}
Jean-Pierre Briot, Ga{\"e}tan Hadjeres, and Fran{\c{c}}ois-David Pachet.
\newblock Deep learning techniques for music generation--a survey.
\newblock {\em arXiv preprint arXiv:1709.01620}, 2017.

\bibitem[\protect\citeauthoryear{Briot \bgroup \em et al.\egroup
  }{2017b}]{briot2017deep}
Jérôme-Pierre Briot, Ghislain Hadjeres, and François Pachet.
\newblock Deep learning for music generation: A survey.
\newblock {\em arXiv preprint arXiv:1709.01620}, 2017.

\bibitem[\protect\citeauthoryear{Buteau and Mazzola}{2008}]{rolland2001motif}
Chantal Buteau and Guerino Mazzola.
\newblock Motivic analysis according to rudolph r{\'e}ti: formalization by a
  topological model.
\newblock {\em Journal of Mathematics and Music}, 2(3):117--134, 2008.

\bibitem[\protect\citeauthoryear{Buteau and
  Vipperman}{2008}]{buteau2001geometrical}
Chantal Buteau and John Vipperman.
\newblock Representations of motivic spaces of a score in openmusic.
\newblock {\em Journal of Mathematics and Music}, 2(2):61--79, 2008.

\bibitem[\protect\citeauthoryear{Cambouropoulos}{1998}]{cambouropoulos1998towards}
Emilios Cambouropoulos.
\newblock Towards a general computational theory of musical structure.
\newblock {\em Technical Report}, 1998.

\bibitem[\protect\citeauthoryear{Collins \bgroup \em et al.\egroup
  }{2010}]{collins2011pattern}
Tom Collins, Jeremy Thurlow, Robin Laney, Alistair Willis, and Paul Garthwaite.
\newblock A comparative evaluation of algorithms for discovering translational
  patterns in baroque keyboard works.
\newblock In {\em Proceedings of the 11th International Society for Music
  Information Retrieval Conference (ISMIR 2010)}, Utrecht, The Netherlands,
  2010.

\bibitem[\protect\citeauthoryear{Conklin and
  Witten}{1995}]{conklin1995multiple}
Darrell Conklin and Ian~H Witten.
\newblock Multiple viewpoint systems for music prediction.
\newblock {\em Journal of New Music Research}, 24(1):51--73, 1995.

\bibitem[\protect\citeauthoryear{Conklin}{2003}]{conklin2003music}
Darrell Conklin.
\newblock Music generation from statistical models.
\newblock In {\em Proceedings of the AISB 2003 Symposium on Artificial
  Intelligence and Creativity in the Arts and Sciences}, pages 30--35.
  Citeseer, 2003.

\bibitem[\protect\citeauthoryear{Copet \bgroup \em et al.\egroup
  }{2024}]{copet2023simple}
Jade Copet, Felix Kreuk, Itai Gat, Tal Remez, David Kant, Gabriel Synnaeve,
  Yossi Adi, and Alexandre D{\'e}fossez.
\newblock Simple and controllable music generation.
\newblock {\em Advances in Neural Information Processing Systems}, 36, 2024.

\bibitem[\protect\citeauthoryear{Deutsch}{2019}]{deutsch2013psychology}
Diana Deutsch.
\newblock Psychology and music.
\newblock In {\em Psychology and its allied disciplines}, pages 155--194.
  Psychology Press, 2019.

\bibitem[\protect\citeauthoryear{Dhariwal \bgroup \em et al.\egroup
  }{2020}]{dhariwal2020jukebox}
Prafulla Dhariwal, Heewoo Jun, Christine Payne, Jong~Wook Kim, Alec Radford,
  and Ilya Sutskever.
\newblock Jukebox: A generative model for music.
\newblock {\em arXiv preprint arXiv:2005.00341}, 2020.

\bibitem[\protect\citeauthoryear{Dong \bgroup \em et al.\egroup
  }{2018}]{dong2018musegan}
Hao-Wen Dong, Wen-Yi Hsiao, Li-Chia Yang, and Yi-Hsuan Yang.
\newblock Musegan: Multi-track sequential generative adversarial networks for
  symbolic music generation and accompaniment.
\newblock In {\em Proceedings of the AAAI Conference on Artificial
  Intelligence}, volume~32, 2018.

\bibitem[\protect\citeauthoryear{Engel \bgroup \em et al.\egroup
  }{2019}]{engel2019gansynth}
Jesse Engel, Kumar~Krishna Agrawal, Shuo Chen, Ishaan Gulrajani, Chris Donahue,
  and Adam Roberts.
\newblock Gansynth: Adversarial neural audio synthesis.
\newblock {\em arXiv preprint arXiv:1902.08710}, 2019.

\bibitem[\protect\citeauthoryear{Guan \bgroup \em et al.\egroup
  }{2018}]{guan2018melodic}
Yixing Guan, Jinyu Zhao, Yiqin Qiu, Zheng Zhang, and Gus Xia.
\newblock Melodic phrase segmentation by deep neural networks.
\newblock {\em arXiv preprint arXiv:1811.05688}, 2018.

\bibitem[\protect\citeauthoryear{Hernandez-Olivan \bgroup \em et al.\egroup
  }{2020}]{hernandez2020music}
Carlos Hernandez-Olivan, Jose~R Beltran, and David Diaz-Guerra.
\newblock Music boundary detection using convolutional neural networks: A
  comparative analysis of combined input features.
\newblock {\em arXiv preprint arXiv:2008.07527}, 2020.

\bibitem[\protect\citeauthoryear{Hong \bgroup \em et al.\egroup
  }{2014}]{hong2014unsupervised}
Yi~Hong, Zhangzhang Si, Wenze Hu, Song-Chun Zhu, and Ying~Nian Wu.
\newblock Unsupervised learning of compositional sparse code for natural image
  representation.
\newblock {\em Quarterly of Applied Mathematics}, pages 373--406, 2014.

\bibitem[\protect\citeauthoryear{Honing}{2006}]{honing2006computational}
Henkjan Honing.
\newblock Computational modeling of music cognition: A case study on model
  selection.
\newblock {\em Music Perception}, 23(5):365--376, 2006.

\bibitem[\protect\citeauthoryear{Janssen \bgroup \em et al.\egroup
  }{2013}]{nieto2016discovering}
Berit Janssen, W~Bas De~Haas, Anja Volk, and Peter Van~Kranenburg.
\newblock Discovering repeated patterns in music: state of knowledge,
  challenges, perspectives.
\newblock In {\em Proc. of the 10th International Symposium on Computer Music
  Multidisciplinary Research, Marseille, France}, volume~20, page~74, 2013.

\bibitem[\protect\citeauthoryear{Lartillot}{2004}]{lartillot2004musical}
Olivier Lartillot.
\newblock A musical pattern discovery system founded on a modeling of listening
  strategies.
\newblock {\em Computer Music Journal}, 28(3):53--67, 2004.

\bibitem[\protect\citeauthoryear{Lattner \bgroup \em et al.\egroup
  }{2015}]{lattner2015probabilistic}
Stefan Lattner, Maarten Grachten, Kat Agres, and Carlos~Eduardo
  Cancino~Chac{\'o}n.
\newblock Probabilistic segmentation of musical sequences using restricted
  boltzmann machines.
\newblock In {\em International Conference on Mathematics and Computation in
  Music}, pages 323--334. Springer, 2015.

\bibitem[\protect\citeauthoryear{Lerdahl and
  Jackendoff}{1996}]{lerdahl1996generative}
Fred Lerdahl and Ray~S Jackendoff.
\newblock {\em A Generative Theory of Tonal Music, reissue, with a new
  preface}.
\newblock MIT press, 1996.

\bibitem[\protect\citeauthoryear{Margrave}{1968}]{margrave1968fundamentals}
Wendell Margrave.
\newblock " fundamentals of musical composition", by a. schoenberg (book
  review).
\newblock {\em Notes}, 25(2):233, 1968.

\bibitem[\protect\citeauthoryear{Margulis}{2013}]{margulis2013repetition}
Elizabeth~Hellmuth Margulis.
\newblock {\em On repeat: How music plays the mind}.
\newblock Oxford University Press, 2013.

\bibitem[\protect\citeauthoryear{Marsden}{2012}]{marsden2010musical}
Alan Marsden.
\newblock Counselling a better relationship between mathematics and musicology.
\newblock {\em Journal of Mathematics and Music}, 6(2):145--153, 2012.

\bibitem[\protect\citeauthoryear{Meredith \bgroup \em et al.\egroup
  }{2002}]{meredith2002algorithms}
David Meredith, Kjell Lemstrom, and Geraint~A Wiggins.
\newblock Algorithms for discovering repeated patterns in multidimensional
  representations of polyphonic music.
\newblock {\em Journal of New Music Research}, 31(4):321--345, 2002.

\bibitem[\protect\citeauthoryear{Meredith}{2015}]{meredith2015music}
David Meredith.
\newblock Music analysis and point-set compression.
\newblock {\em Journal of New Music Research}, 44(3):245--270, 2015.

\bibitem[\protect\citeauthoryear{Meyer}{2008}]{meyer2008emotion}
Leonard~B Meyer.
\newblock {\em Emotion and meaning in music}.
\newblock University of chicago Press, 2008.

\bibitem[\protect\citeauthoryear{M{\"u}ller}{2015}]{muller2015fundamentals}
Meinard M{\"u}ller.
\newblock {\em Fundamentals of music processing: Audio, analysis, algorithms,
  applications}, volume~5.
\newblock Springer, 2015.

\bibitem[\protect\citeauthoryear{Perkins}{2022}]{perkins2022musical}
Reed~James Perkins.
\newblock Musical phrase segmentation via grammatical induction.
\newblock Master's thesis, Brigham Young University, 2022.

\bibitem[\protect\citeauthoryear{{Riffusion}}{2023}]{riffusion2023}
{Riffusion}.
\newblock Riffusion: Ai music generation.
\newblock \url{https://riffusion.com/}, 2023.
\newblock Accessed: 2025-02-11.

\bibitem[\protect\citeauthoryear{Simoni and
  Dannenberg}{2013}]{collins2011algorithmic}
Mary Simoni and Roger~B Dannenberg.
\newblock {\em Algorithmic composition: a guide to composing music with
  nyquist}.
\newblock University of Michigan Press, 2013.

\bibitem[\protect\citeauthoryear{Van~Geert and Wagemans}{2020}]{van2020order}
Eline Van~Geert and Johan Wagemans.
\newblock Order, complexity, and aesthetic appreciation.
\newblock {\em Psychology of aesthetics, creativity, and the arts}, 14(2):135,
  2020.

\bibitem[\protect\citeauthoryear{Wang \bgroup \em et al.\egroup
  }{2015}]{li2007music}
Cheng-i Wang, Jennifer Hsu, and Shlomo Dubnov.
\newblock Music pattern discovery with variable markov oracle: A unified
  approach to symbolic and audio representations.
\newblock In {\em ISMIR}, pages 176--182, 2015.

\bibitem[\protect\citeauthoryear{Wu \bgroup \em et al.\egroup
  }{2010}]{wu2010learning}
Ying~Nian Wu, Zhangzhang Si, Haifeng Gong, and Song-Chun Zhu.
\newblock Learning active basis model for object detection and recognition.
\newblock {\em International journal of computer vision}, 90:198--235, 2010.

\bibitem[\protect\citeauthoryear{Zipf}{2016}]{zipf2016human}
George~Kingsley Zipf.
\newblock {\em Human behavior and the principle of least effort: An
  introduction to human ecology}.
\newblock Ravenio books, 2016.

\end{thebibliography}

\end{document}